\documentclass{article}
\usepackage{spconf,amsmath,graphicx}

\usepackage{mwe} % to get dummy images

\usepackage[hidelinks]{hyperref}
\usepackage{url}

\usepackage{graphicx} % Required for including images
\graphicspath{{./Figures/}}
\usepackage{mathtools,amsmath,amssymb,amsthm} % For including math equations, theorems, symbols, etc
\usepackage[overload]{empheq}
\usepackage{enumitem}
\usepackage{multicol}
\usepackage{capt-of}
\usepackage[export]{adjustbox}

\theoremstyle{remark}

\usepackage{siunitx}
\usepackage{booktabs}
\usepackage{xcolor}
\usepackage[ruled,vlined]{algorithm2e}
\usepackage{breqn}
\usepackage{multirow}
\usepackage{mathrsfs}
\usepackage{breqn}
\usepackage{mathtools}
\usepackage{setspace}

\DeclareMathAlphabet\mathbfcal{OMS}{cmsy}{b}{n}

\title{Fast High Resolution Blood Flow Estimation and Clutter Rejection via an Alternating Optimization Problem}
\author{Duong-Hung~Pham, Adrian Basarab, Ilyess Zemmoura, Paul Rodríguez, Jean-Pierre Remenieras and  Denis Kouam\'e}
\name{Duong-Hung~Pham$^{1}$, Adrian Basarab$^{1}$, Jean-Pierre Remenieras$^{2}$, Paul Rodríguez$^{3}$, Denis Kouam\'e$^{1}$}
\address{$^{1}$IRIT, CNRS UMR 5505, Paul Sabatier University, Toulouse 31062, France \\
         $^{2}$UMR 1253, iBrain, Tours University, Tours 37032, France\\
         $^{3}$Pontifical Catholic University of Peru, Av. Universitaria 1801, San Miguel, Lima 32, Perú}
     
\let\OLDthebibliography\thebibliography
\renewcommand\thebibliography[1]{
  \OLDthebibliography{#1}
  \setlength{\parskip}{0pt}
  \setlength{\itemsep}{0pt plus 0.3ex}
}     

\begin{document}
%\ninept
%
\maketitle
\begin{abstract}
This paper introduces a computationally efficient technique for estimating high-resolution Doppler blood flow from an ultrafast ultrasound image sequence. More precisely, it consists in a new fast alternating minimization algorithm that implements a blind deconvolution method based on robust principal component analysis. Numerical investigation carried out on \textit{in vivo} data shows the efficiency of the proposed approach in comparison with state-of-the-art methods. 
\end{abstract}
\section{Introduction}
\vspace{-0.2cm}
\label{sec:intro}
Estimating high-sensitivity and high-resolution blood flow from an ultrafast ultrasound (US) sequence has received a great attention from the medical imaging community over the last few years. The main objective is to provide a clear and accurate 3D visualization of vascular networks inside the human body and consequently lead to a better prognosis and treatment of many related diseases such as brain gliomas (tumors) or peri-tumoral infiltration. To this end, the spatio-temporal singular value decomposition (SVD)  of the dataset Casorati matrix, able to suppress temporally clutter signals originating from quasi-static tissues from blood flow, is the most widespread method \cite{demene_spatiotemporal_2015}. Unfortunately, the manual choice of the correlation thresholds for separating the blood space from the tissue and noise subspaces partially mitigates the applicability of this method. To overcome this drawback, many efforts have been made in the literature, among which the most popular relies on the robust principal component analysis (RPCA) \cite{Wright_robust_2009}, or its variants, e.g, \cite{Fatemi2018}. Moreover, to account for the loss of spatial resolution of the Doppler data caused by the system point spread function (PSF), recently a deconvolution step was embedded in RPCA-based methods, allowing high-resolution and high-sensitivity blood flow estimation \cite{Shen2019}. This method requires the knowledge of the PSF, that can be experimentally measured or estimated jointly with the blood flow, as suggested by the recent blind deconvolution (BD) approach, called BD-RPCA in \cite{Pham2020}. Despite its accuracy, BD-RPCA suffers from a high computational cost. 

In this paper, we propose a new computationally efficient algorithm in which the nuclear norm term related to the BD-RPCA model is replaced by a fixed rank constraint, followed by a fast partial SVD-based alternating algorithm. The remainder of the paper is organized as follows. The background about the model and related works is given in Section \ref{sec:estab}. The proposed fast blind algorithm of blood flow estimation is detailed in Section \ref{sec:fBDRPCA}. Finally, numerical results on \textit{in vivo} ultrafast US data are regrouped in Section \ref{sec:Results}, showing the improvement achieved by the proposed approach over some existing techniques.  

%%%%%%%%Model establishment %%%%%%%%%%
\section{Background}\label{sec:estab}
%\vspace{-0.2cm}
\subsection{Model formulation}
\vspace{-0.2cm}
\label{subsec:RPCA}
Let us consider a temporal in-phase and quadrature (IQ) Doppler ultrafast US sequence, composed of $N_t$ frames of size $ N_x \times N_z$, with depth $N_z$, probe width $N_x$ and acquisition time $N_t$. $\boldsymbol{S} \in \mathbb{C}^{N_z N_x\times N_t}$ denotes the related Casorati matrix. The tissue-blood flow model is written as follows \cite{demene_spatiotemporal_2015}: 
\setlength{\abovedisplayskip}{0pt}
\setlength{\belowdisplayskip}{0pt}
\begin{equation}
\boldsymbol{S} = \boldsymbol{T}+\boldsymbol{B}+\boldsymbol{N},
\label{eq:basic}
\end{equation}
\noindent where $\boldsymbol{T}, \boldsymbol{B}, \boldsymbol{N}  \in \mathbb{C}^{N_z N_x\times N_t}$ are the tissue, blood and additive noise matrices, respectively. To avoid further notations, the vectorized counterparts of these matrices are, when needed, considered hereafter and defined using the same notations. In most practical applications, the blood flow $\boldsymbol{B}$ is assumed to be sparse, conventionally promoted by the $l_1$-norm, while the tissue $\boldsymbol{T}$ possesses a slight change over time, usually modelled by the nuclear norm $||.||_*$. Also, $\boldsymbol{N}$ is assumed to be an additive Gaussian noise. The main objective of this work is to recover $\boldsymbol{B}$ and $\boldsymbol{T}$ from $\boldsymbol{S}$ based on building an optimization problem imposing appropriate constraints on these matrices.
%\vspace{-0.4cm}
\subsection{Blind deconvolved RPCA}
\vspace{-0.2cm}
\label{sec:BD-RPCA}
An appealing method, called blind deconvolved robust principal component analysis (BD-RPCA), has been recently proposed in \cite{Pham2020}. It aims at reconstructing a high resolution blood flow $\boldsymbol{X}$ simultaneously with the tissue component $\boldsymbol{T}$ and the system PSF $\mathbfcal{H}_e$ from the acquired US Doppler data. The associated optimization problem is formulated as follows:
\begin{align} 
\label{eq:bdrpca} 
    [\hat{\boldsymbol{X}},\hat{\mathbfcal{H}_e},\hat{\boldsymbol{T}}] &= \mathop{\arg\min_{{\boldsymbol{X}},{\mathbfcal{H}_e},{\boldsymbol{T}}}}  \left\{ ||\boldsymbol{S}-\mathbfcal{H}_e\circledast \boldsymbol{X}-\boldsymbol{T}||^2_F  \right. \notag\\
      & + \left.  \lambda||\boldsymbol{X}||_1 + \rho||\boldsymbol{T}||_*  \right\}, \text{s.t.~~} |\mathscr{F}(\mathbfcal{H}_e)| = \tilde{\textbf{H}}, 
\end{align}
\noindent where $\circledast$ denotes the 2D convolution, $\hat{.}$ the estimated variables, $||.||^2_F$ the Frobenius norm and $\lambda, \rho>0$ two hyper-parameters balancing the trade-off between the sparsity of the blood and the low-rankness of the tissue \cite{Wright_robust_2009,Fatemi2018}. $\tilde{\textbf{H}}$ is the magnitude of the 2D Fourier transform ($\mathscr{F}$) of the PSF that can be straightforwardly computed from $\boldsymbol{S}-\hat{\boldsymbol{T}}$ using homomorphic filtering \cite{Taxt1995}. For computational efficiency, the above 2D convolution, supposed circulant, is rewritten as an element-wise multiplication in the Fourier domain. To solve \eqref{eq:bdrpca}, a two-step alternating algorithm was proposed in \cite{Pham2020}, resumed hereafter. 
\begin{enumerate}[label=\roman*),leftmargin=0.3cm,itemsep=0mm]
\item  For fixed $\boldsymbol{X}$ and $\boldsymbol{T}$ and using spatio-temporally invariant assumption on PSF $\mathbfcal{H}_e$, \eqref{eq:bdrpca} can be reformulated as: 
\begin{align}
&[\hat{\mathbfcal{H}_e}^{(k+1)}] =
\mathop{\arg\min_{{\mathbfcal{H}_e}}} \left\{ ||\overline{\sum\limits_{N_t}}\left(\boldsymbol{S}-\boldsymbol{T}^{(k+1)}\right) \right. \notag \\
      &\hspace{0.1cm}- \left. \mathbfcal{H}_e\circledast \overline{\sum\limits_{N_t}}(\boldsymbol{X}^{(k+1)})||^2_F \right\}, \text{s.t.} |\mathscr{F}(\mathbfcal{H}_e)| = \tilde{\textbf{H}},
\label{eq:bdrpca2}
\end{align} 
\noindent where $\overline{\sum\limits_{N_t}}(\boldsymbol{Z})$ stands for the temporal average of a 2D matrix $ \boldsymbol{Z}$ obtained by taking the mean along the time dimension of the 3D version of this matrix. Then, (\ref{eq:bdrpca2}) can be solved by the blind deconvolution (BD) technique \cite{Michailovich2019}. \item For a fixed $\mathbfcal{H}_e$, \eqref{eq:bdrpca} reads as:
{\small
\begin{align} 
\label{eq:drpca}
    [\hat{\boldsymbol{X}}^{(k+1)},\hat{\boldsymbol{T}}^{(k+1)}]  &= \mathop{\arg\min_{{\boldsymbol{X}},{\boldsymbol{T}}}}  \left\{ ||\boldsymbol{S}-\mathbfcal{H}_e^{(k+1)} \circledast \boldsymbol{X}-\boldsymbol{T}||^2_F  \right. \notag \\
      &\hspace{1.3cm} + \left.  \lambda||\boldsymbol{X}||_1 + \rho||\boldsymbol{T}||_*  \right\}. 
\end{align}
}
To solve \eqref{eq:drpca}, the alternating direction method of multipliers (ADMM) was used \cite{Shen2019,Boyd_distributed_2010}. It consists in performing, at each iteration $(k+1)$, the five following main steps:
{\small
\begin{align}
 &\hat{\boldsymbol{T}}^{(k+1)}=\mathop{\arg\min_{\boldsymbol{T}}}\left\{\rho||\boldsymbol{T}||_*\right. \notag\\ &\hspace{1.0cm} \left. +\frac{\mu}{2}||\boldsymbol{T} 
      -(\boldsymbol{S}-\mathbfcal{H}_e^{(k+1)} \circledast\boldsymbol{X}^{(k)}+\frac{1}{\mu} \boldsymbol{\nu}^{(k)})||^2_F\right\}  \label{eq:algo_bdrpca} , \\
 & \hat{\boldsymbol{z}}^{(k+1)}=\mathop{\arg\min_{\boldsymbol{z}}}\left\{\lambda||\boldsymbol{z}||_1+\frac{\mu}{2}||\boldsymbol{z}-(\boldsymbol{X}^{(k)}+\frac{1}{\mu} \boldsymbol{w}^{(k)})||^2_F\right\},  
 \end{align}
 }
 {\small
\begin{align}
   &\hat{\boldsymbol{X}}^{(k+1)}=\mathop{\arg\min_{\boldsymbol{X}}}\left\{\frac{\mu}{2}||\boldsymbol{S}-\mathbfcal{H}_e^{(k+1)} \circledast\boldsymbol{X} - \boldsymbol{T}^{(k+1)}  \right. \notag\\
      &\hspace{1.5cm} \left. + \frac{1}{\mu} \boldsymbol{\nu}^{(k)})||^2_F +\frac{\mu}{2}||\boldsymbol{X}-\boldsymbol{z}^{(k+1)}+\frac{1}{\mu} \boldsymbol{w}^{(k)}||^2_F\right\},  \\
  &\hat{\boldsymbol{\nu}}^{(k+1)}=\boldsymbol{\nu}^k+\mu(\boldsymbol{S}-\mathbfcal{H}_e^{(k+1)} \circledast\boldsymbol{X}^{(k+1)}-\boldsymbol{T}^{(k+1)}), \\
  &\hat{\boldsymbol{w}}^{(k+1)}=\boldsymbol{\nu}^{(k)}+\mu (\boldsymbol{X}^{(k+1)}-\boldsymbol{z}^{(k+1)}),
\end{align}
}
\noindent where $\boldsymbol{z}$ is an auxiliary variable equal to $\boldsymbol{X}$, $\boldsymbol{\nu}$ the Lagrange multiplier and $\mu$ the Lagrangian penalty parameter. 
\end{enumerate} 

Although BD-RPCA has been shown to be an efficient method for estimating high-resolution blood flow together with the tissue and the PSF, it is not computationally attractive. Indeed, the estimation of $\hat{\boldsymbol{T}}^{(k+1)}$ in \eqref{eq:algo_bdrpca}, is a convex problem that has an analytical solution corresponding to a singular value thresholding (SVT) \cite{Cai2010}. This step is computationally expensive since it requires a full SVD decomposition of a large filtered matrix, of the size of the acquired spatio-temporal data. 

\section{Proposed fast BD-RPCA method} \label{sec:fBDRPCA}
\vspace{-0.2cm}
The main aim of the proposed algorithm is to reduce the computational time of BD-RPCA by introducing a fast variant of \eqref{eq:drpca} and subsequently combining it with BD as explained in subsection \ref{sec:BD-RPCA}. More precisely, instead of imposing the nuclear norm relaxation in \eqref{eq:drpca}, the rank is itself constrained as follows:
{\small
\begin{align} 
\label{eq:fbdrpca} 
 &[\hat{\boldsymbol{X}}^{(k+1)},\hat{\boldsymbol{T}}^{(k+1)}]  = \mathop{\arg\min_{{\boldsymbol{X}},{\boldsymbol{T}}}}  \left\{ ||\boldsymbol{S}-\mathbfcal{H}_e^{(k+1)} \circledast \boldsymbol{X}-\boldsymbol{T}||^2_F  \right. \notag \\
       & \hspace{2.0cm}+ \left.  \lambda||\boldsymbol{X}||_1  \right\} \text{~~s.t.~}  \text{rank}(\boldsymbol{T}) = r_f,
\end{align}
}
\noindent where $\mathbfcal{H}_e^{(k+1)}$ is assumed to be known from the previous iteration and $r_f$ is a rank hyperparameter that needs to be tuned. Interestingly, the reference value of $r_f$ can be efficiently estimated using a two-step algorithm called \emph{Rank Estimation} developed for low-rank matrix completion problem in \cite{Keshavan2010}. The idea behind this technique is to find the index $i$ corresponding to the rank value that minimizes the following cost function:
\begin{align}
    R(i) = \frac{\sigma_{i+1} + \sigma_ {1} \sqrt{\frac{i}{\zeta}}}{\sigma_i}, 
\end{align}
where $\zeta = \frac{|\mathbfcal{C}_{\boldsymbol{S}}|}{\sqrt{N_z N_x\times N_t}}$ with $|\mathbfcal{C}_{\boldsymbol{S}}|$ and $\sigma_{i}$ respectively the cardinality, and the singular values of the acquired Casorati matrix.
 
To solve (\ref{eq:fbdrpca}), we propose herein a two-step alternating minimization algorithm as follows:
{\small
\begin{align} 
    &\hat{\boldsymbol{T}}^{(k+1)}= \mathop{\arg\min_{\boldsymbol{T}}}  \left\{ ||\left(\boldsymbol{S}-\mathbfcal{H}_e^{(k+1)}\circledast \boldsymbol{X}^{(k)}\right)-\boldsymbol{T}||^2_F \right\} \notag\\
    & \hspace{4cm} \text{~~s.t.~}  \text{rank}(\boldsymbol{T}) = r_f, \label{eq:fbdrpca1} \\
    &\hat{\boldsymbol{X}}^{(k+1)} = \mathop{\arg\min_{\boldsymbol{X}}}  \left\{ ||\left(\boldsymbol{S}-\boldsymbol{T}^{(k+1)}\right)-\mathbfcal{H}_e^{(k+1)}\circledast \boldsymbol{X}||^2_F  \right. \notag\\
      & \hspace{4.5cm}+ \left.  \lambda||\boldsymbol{X}||_1  \right\}. \label{eq:fbdrpca2}  
\end{align}
}
Subproblem (\ref{eq:fbdrpca1}) can be rapidly solved based on computing a partial SVD decomposition with only the first $r_f$ components of $\boldsymbol{S}-\mathbfcal{H}_e^{(k+1)}\circledast \boldsymbol{X}^{(k)}$ \cite{Rodriguez2013}. In particular, for small $r_f$, this decomposition can be performed consistently faster when using the Lanczos bidiagonalization algorithm with partial reorthogonalization through \emph{lansvd} routine from PROPACK library \cite{Larsen1998,Larsen2005}. Subproblem (\ref{eq:fbdrpca2}) consisting in a well-known $l_1$ regularized least absolute shrinkage and selection operator (LASSO) regression can be simply solved efficiently by means of an ADMM-based algorithm \cite{Boyd_distributed_2010}.

The pseudo algorithm associated with the proposed method, called fast BD-RPCA, is reported in Algorithm \ref{alg:fdrpca}. The initial values of the blood and tissue are guessed using SVD for a more efficient convergence of the algorithm \cite{Pham2020}. 
\begin{algorithm}
 \label{alg:fdrpca}
\SetAlgoLined
\KwIn{Casorati matrix $\boldsymbol{S}$}
\textbf{Initialize:} $\varepsilon=10^{-6}$,  $[\boldsymbol{X}^{(0)},\boldsymbol{T}^{(0)}]$ = \text{SVD}($\boldsymbol{S}$)\\
 \While{$||\boldsymbol{X}^{(k+1)}-\boldsymbol{X}^{(k)}||_F > \varepsilon$}{  
   \begin{enumerate}[leftmargin=0.2cm]
     \setlength\itemsep{0em}
    \item compute temporal average: $\boldsymbol{M}_{\boldsymbol{ST}}^{(k+1)}=\overline{\sum\limits_{N_t}}\left(\boldsymbol{S}-\boldsymbol{T}^{(k)}\right)$ 
   	\item estimate PSF: $\mathbfcal{H}_{e}^{(k+1)}$ = \text{BD}$\left(\boldsymbol{M}_{\boldsymbol{ST}}^{(k+1)}\right)$, \d
   	using Eq. (\ref{eq:bdrpca2})
    \item update $[\boldsymbol{X}^{(k+1)},\boldsymbol{T}^{(k+1)}]$ using new fast procedure: \begin{enumerate}[leftmargin=0.3cm]
     \setlength\itemsep{0em}
    \item compute a partial SVD and $\boldsymbol{T}^{(k+1)}$:
    $[U, \Delta, V]=\text{SVD}\left( \boldsymbol{S}-\mathbfcal{H}_e^{(k+1)}\circledast \boldsymbol{X}^{(k)}, r_f\right)$,\\
    $\boldsymbol{T}^{(k+1)} = U\Delta V^{\text{T}}$
    \item estimate $\boldsymbol{X}^{(k+1)} =\text{LASSO} \left(\boldsymbol{S}-\boldsymbol{T}^{(k+1)}, \mathbfcal{H}_e^{(k+1)}\right) $ using Eq. (\ref{eq:fbdrpca2}) \vspace{-0.5cm}
    \end{enumerate}  
    \end{enumerate}}     
 \KwOut{$\boldsymbol{X}^{(k+1)}$, $\boldsymbol{T}^{(k+1)}$ and $\mathbfcal{H}_{e}^{(k+1)}$}
 \caption{fast BD-RPCA}
\end{algorithm}
%%%%% Results %%%%%%%%%%%%%%%
\section{Numerical Results} \label{sec:Results}
\vspace{-0.2cm}
This section regroups numerical results on \textit{in vivo} US data to demonstrate the improvement achieved by the proposed approach over other existing methods including SVD \cite{demene_spatiotemporal_2015}, randomized low-rank \& sparse matrix decomposition (GoDec) \cite{Tianyi2011} and BD-RPCA \cite{Pham2020}. All the experiments were conducted using MATLAB R2019b on a computer with Intel(R) Core(TM) i5-8500 CPU @3.00 GHz and 16GB RAM. \footnote{The Matlab scripts implementing the method and generating all figures are available at \url{github.com/phamduonghung/fast_BDRPCA}}
\vspace{-0.2cm}
\subsection{Data acquisition}
\vspace{-0.2cm}
The ultrafast Doppler data was acquired in Regional University Hospital Bretonneaux of Tours – Department of Neurosurgery from a patient undergoing brain surgery with open skull whose dura mater had been removed. The AixplorerTM (Supersonic Imagine) ultrasound scanner with the SL10-2 probe (192 elements) was used for US acquisition. The research package (SonicLab V12) was used to obtain a particular US sequence of $1000$ frames, compounded angles $[- 5\textsuperscript{{o}}, 0\textsuperscript{{o}}, +5\textsuperscript{{o}}]$ with pulse repetition frequency PRF=3KHz, frame rate 1KHz, imaging depth [1mm-40mm]. The resultant dataset size was $260\times192\times1000$ pixels.
\begin{figure*}[!htb]
\begin{center}
\begin{minipage}[b]{.23\linewidth}
\centering
\centerline{\includegraphics[width=0.95\textwidth, height = 4.0cm]{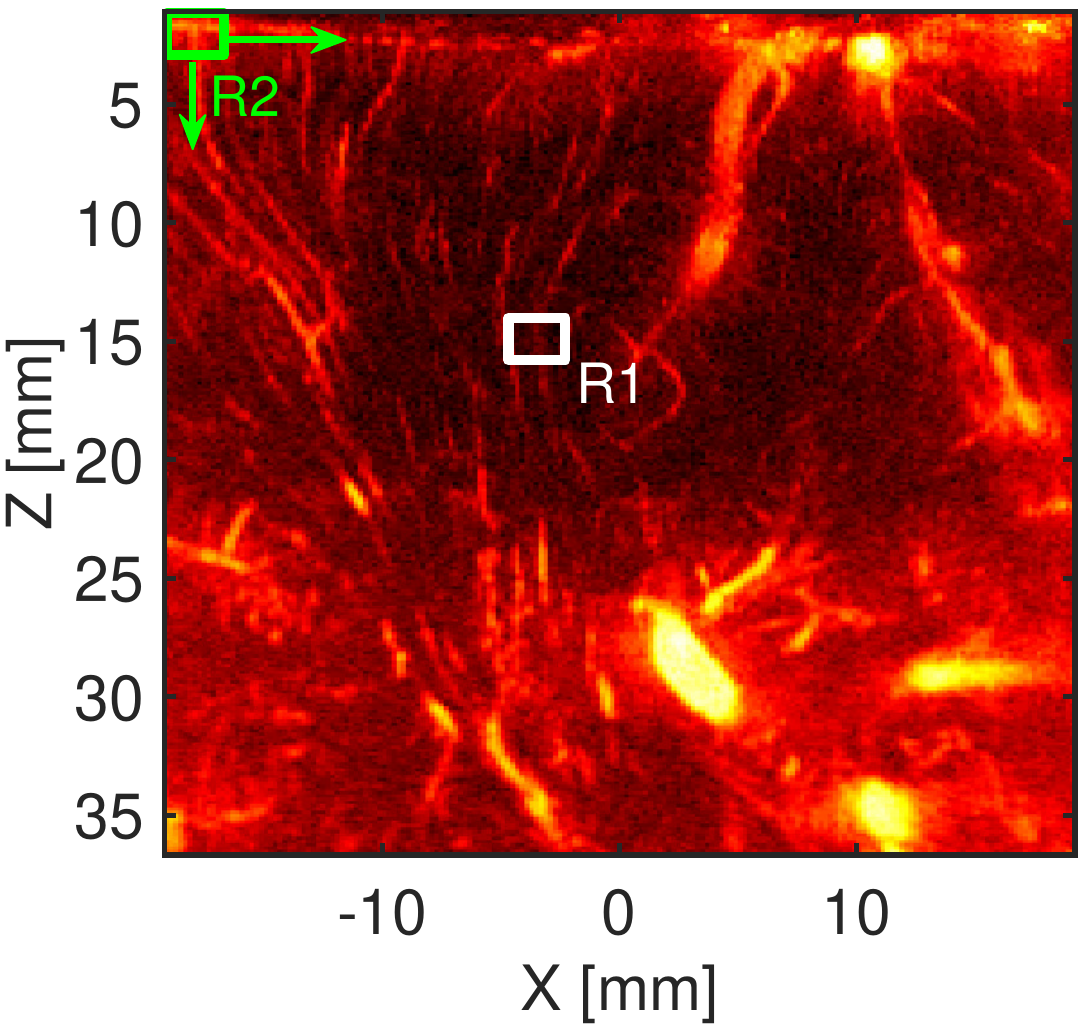}}
\centerline{(a)} 
%\centerline{{\scriptsize(a) RPCA, \emph{CR=-20.96dB}}} 
\end{minipage}
\begin{minipage}[b]{0.23\linewidth}
\centering
\centerline{\includegraphics[width=0.95\textwidth, height = 4.0cm]{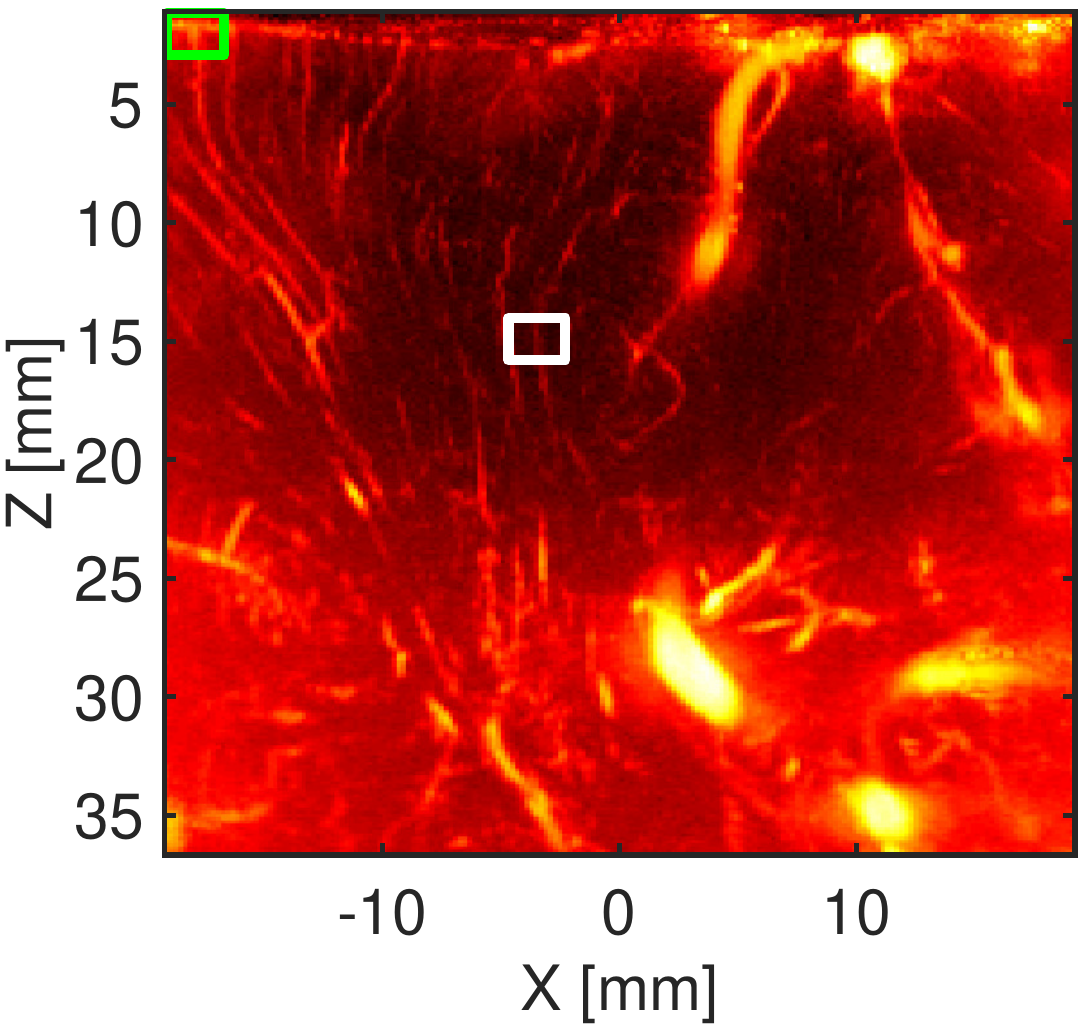}}
\centerline{(b)} 
\end{minipage} 
\begin{minipage}[b]{0.23\linewidth}
\centering
\centerline{\includegraphics[width=0.95\textwidth, height = 4.0cm]{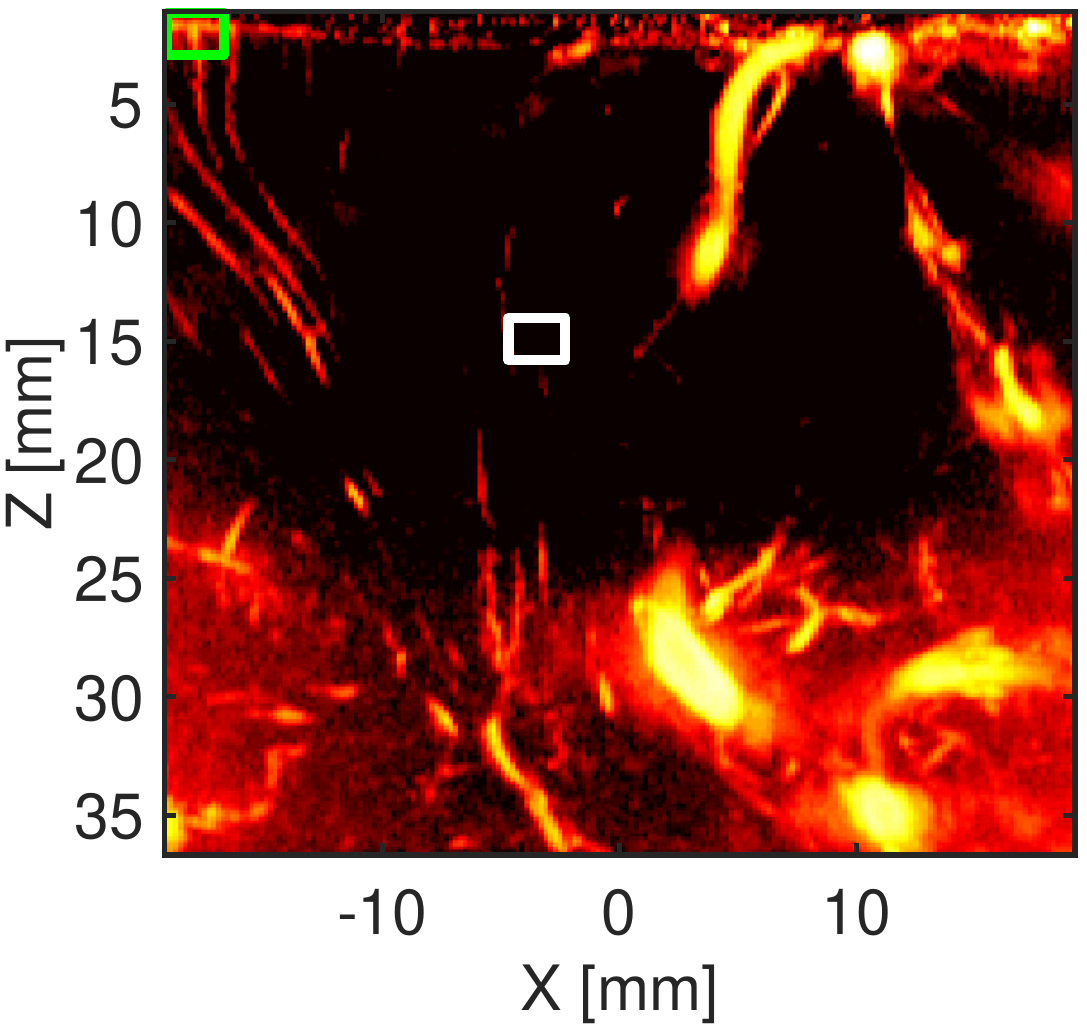}}
\centerline{ (c)} 
\end{minipage} 
\begin{minipage}[b]{0.23\linewidth}
\centering
\centerline{\includegraphics[width=0.95\textwidth, height = 4.0cm]{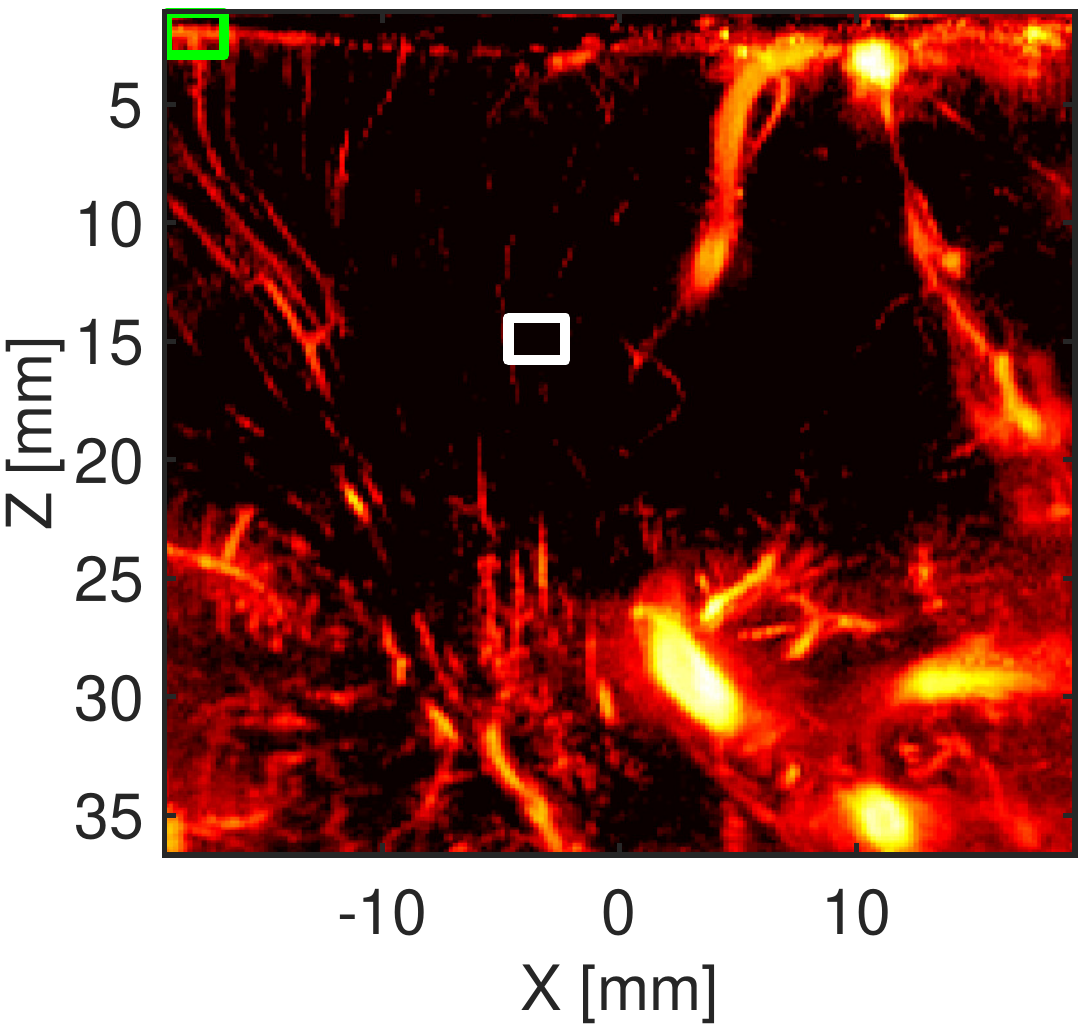}}
\centerline{(d)} 
\end{minipage}
\begin{minipage}[b]{0.05\linewidth}
\centering
\centerline{\includegraphics[trim={11.2cm 0.03cm 0 0},clip,width=0.7\textwidth, height = 4.07cm]{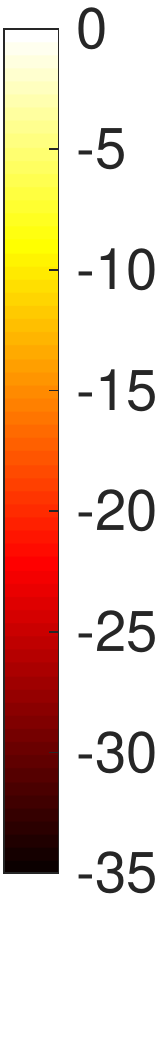}}
%%\vspace{-0.2cm}
\centerline{} 
\end{minipage}
\end{center}
\vspace{-0.7cm}
\caption{Power Doppler images obtained by (a): SVD; (b): GoDec; (c) BD-RPCA; (d): fast BD-RPC.}    
\label{fig:vivo}
\vspace{-0.3cm}
\end{figure*}

\begin{figure}[!htb]
\begin{minipage}[b]{\linewidth}
\centering
\centerline{\includegraphics[width=0.8\textwidth, height = 4.2cm]{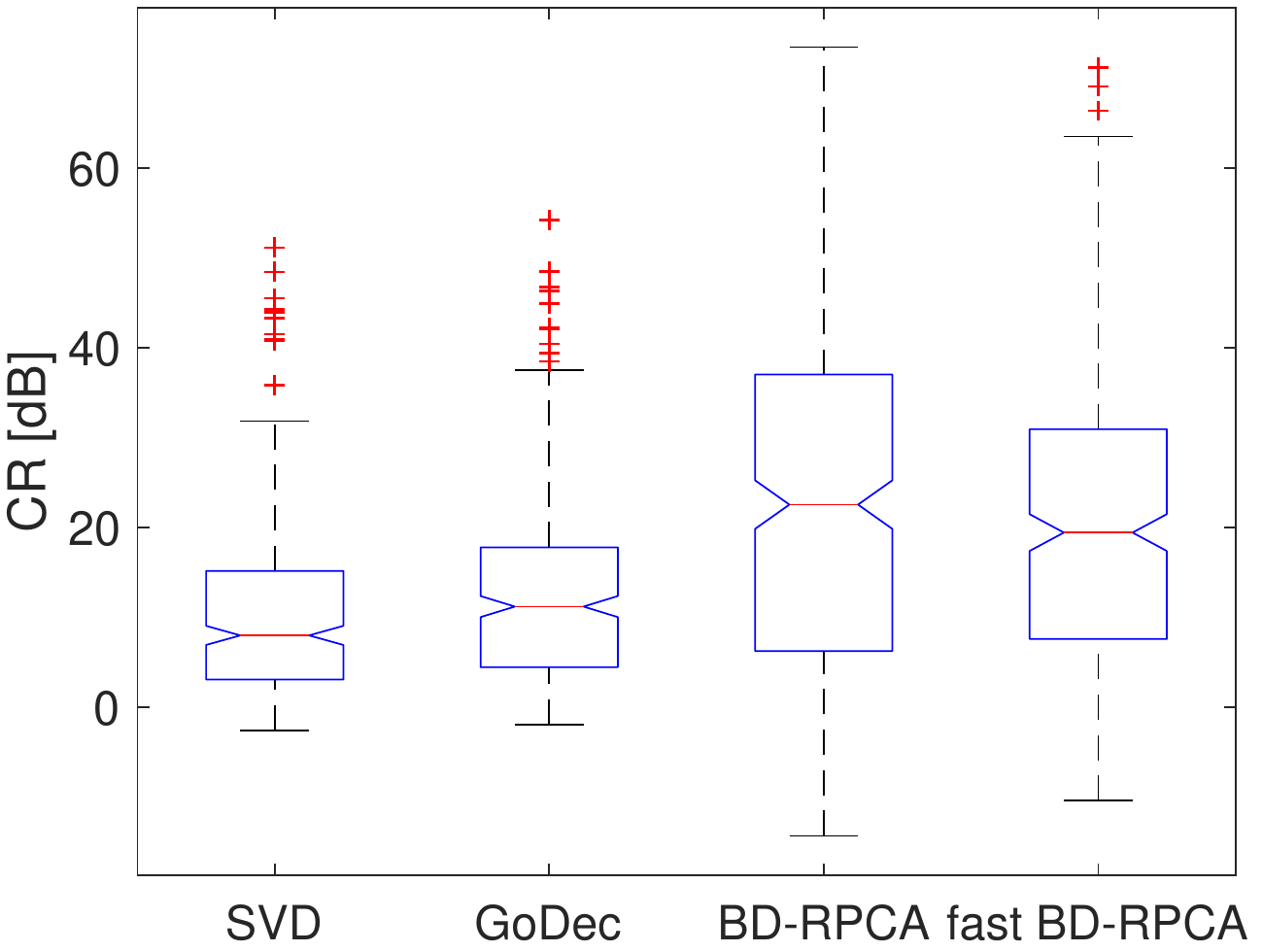}}
\end{minipage}
\vspace{-0.5cm}
\caption{CR measurements for the different tested methods. 
%. In these boxplots, the red horizontal line indicates the median, the bottom and top edges of the box indicating the 25th and 75th percentiles and the black lines indicating the entire range of data samples, per category. Red markers indicate outliers which are excluded from the statistical calculations \cite{McGill1978}.
} \label{fig:CR}
\vspace{-0.5cm}
\end{figure}
\vspace{-0.2cm}
\subsection{Performance comparison}
\vspace{-0.2cm}
To ensure fair comparison, the hyperparameters corresponding to each tested method were tuned by cross-validation to the best possible values. Table \ref{tab:lambd_mu} regroups the optimal hyperparameters associated with SVD, GoDec, BD-RPCA, and fast BD-RPCA. Moreover, for visualization purpose, the Power Doppler image (in dB) was used and defined from the estimated blood flow $\boldsymbol{X}$, for a given position $(x,z)$, as follows:

\begin{equation*}
\boldsymbol{I}_{PD}(x,z) = 10\log10\left(\frac{1}{N_t}\sum\limits_{k=1}^{N_t}\boldsymbol{X}(x,z,k)^{2}\right).
\end{equation*}  
\vspace{-0.5cm}
\begin{table}[!htb]
\centering
\renewcommand{\arraystretch}{1.2}
\caption{Optimal hyperparameter setting}
\label{tab:lambd_mu}
\begin{tabular}{c|c|}
\hline
\multicolumn{1}{|l|}{\bfseries SVD}   &  $T_c = 100$, $T_b = 150$ \\ \hline
\multicolumn{1}{|l|}{\bfseries GoDec }  & $r_G=41, \tau_{G} = 5\times 10^3$  \\ \hline
\multicolumn{1}{|l|}{\bfseries BD-RPCA }  & $\lambda_{B} = 0.0051, \mu_{B} = 0.0403, \rho_{B} = 1$ \\ \hline
\multicolumn{1}{|l|}{\bfseries  fast BD-RPCA }  & $\lambda_{f} = 8\times 10^{-5}, r_f = 41$ \\ \hline
\end{tabular}
\end{table} 
\vspace{-0.2cm}

In Fig.~\ref{fig:vivo}, we depict the Power Doppler image results obtained by the four methods carried out on peritumoral data. Visually examining these plots, one may remark that the Power Doppler images estimated by both SVD and GoDec are very noisy and blurred while those given by BD-RPCA and fast BD-RPCA, overall, provide clear and similar pictures of vessel structures with high resolution. To quantitatively compare these methods, contrast ratio (CR) was used \cite{RodriguezMolares2018}:
\begin{equation*}
\text{CR}_{[\text{dB}]} = 20\log10 \left(\frac{\mu_{\text{R}_2}}{\mu_{\text{R}_1}}\right),
\end{equation*} 
where $\mu_{R_1}$ (resp. $\mu_{R_2}$) is the mean value of intensities in the white (resp. green) rectangular patch $\text{R}_1$ (resp. $\text{R}_2$). Furthermore, $\text{R}_1$ is fixed while $\text{R}_2$ is moved around inside each Power Doppler image so that the resulting CR values are represented by a boxplot as in Fig.~\ref{fig:CR}. From this boxplot, one can notice that the fast proposed algorithm exhibits an equivalent result to BD-RPCA, and outperforms SVD and GoDec.

Moreover, Table 2 shows the running time corresponding to each studied method on the peritumoral dataset. It is noticeable that as expected, the computational times obtained by both SVD and GoDec are the lowest while the one given by the fast proposed algorithm is significantly lower than BD-RPCA. A much lower computational time of SVD or GoDec stems from the fact the the blind deconvolution approach is at the expense of much more running time because of the algorithmic complexity, but ensures considerably higher resolution results. 

To conclude, the above numerical \textit{in vivo} results plead in favour of using fast BD-RPCA, not requiring PSF measurements and well balancing the trade-off between the low computational cost and the high quality blood flow reconstruction, compared to the other studied methods.     
\vspace{-0.5cm}
\begin{table}[!htb]
\centering
\label{tab:running_time1}
\caption{Running times in s for each method}
\begin{tabular}{|c|c|c|c|}
\hline
\textbf{SVD}      & \textbf{GoDec}      & \textbf{BD-RPCA}      & \textbf{fast BD-RPCA}      \\ \hline
   10.38 & 3.7 &  219.9 &  89.2 \\ \hline
\end{tabular}%
\end{table}
\vspace{-0.45cm}

%%%%%%%%%%%%%%%%%%%%%%%%%% Conclusion %%%%%%%%%%%%%%%%%%%%%%%%%%%%%%%
\section{Conclusion}
\label{sec:concl}
\vspace{-0.25cm}
In this paper, we introduced a novel fast algorithm for estimating the blood flow and the tissue from an ultrafast sequence of US images, based on improving the model of BD-RPCA. The proposed technique not only enabled to handle the limitation related to the high computational cost of the latter but also providing similar recovery performances. Numerical simulations carried out on \textit{in vivo} peritumoral data showed the the benefits of using the proposed algorithm. Future work could be devoted, for instance, to evaluate the clinical contribution of the proposed method in the prognosis and treatment of blood-related diseases, or to extend the proposed method to a spatio-temporally invariant PSF model.                 
%\vspace{-0.1cm}

% -------------------------------------------------------------------------
%\newcommand{\BIBdecl}{\setlength{\itemsep}{-0.1em}}
%\vspace{-0.05cm}
\section{Compliance with Ethical Standards}
\label{sec:ethics}
\vspace{-0.3cm}
The data acquisition was done with a clinical research protocol (ELASTOGLI) approved by the institutional review board (CCP: ‘Comité de Protection des Personnes’, $N\textsuperscript{{o}}$ 123748) and local ethical committee. It strictly complies with the ethical principles for medical research involving human subjects of the World Medical Association Declaration of Helsinki.
\vspace{-0.07cm}
\section{Acknowledgments}
\label{sec:acknowledgments}
\vspace{-0.35cm}
The authors declare no conflict of interest, and  non financial interests to disclose. 
\vspace{-0.25cm}
\bibliographystyle{IEEEtran}
\bibliography{refs_ISBI20}

\begin{thebibliography}{10}
\providecommand{\url}[1]{#1}
\csname url@samestyle\endcsname
\providecommand{\newblock}{\relax}
\providecommand{\bibinfo}[2]{#2}
\providecommand{\BIBentrySTDinterwordspacing}{\spaceskip=0pt\relax}
\providecommand{\BIBentryALTinterwordstretchfactor}{4}
\providecommand{\BIBentryALTinterwordspacing}{\spaceskip=\fontdimen2\font plus
\BIBentryALTinterwordstretchfactor\fontdimen3\font minus
  \fontdimen4\font\relax}
\providecommand{\BIBforeignlanguage}[2]{{%
\expandafter\ifx\csname l@#1\endcsname\relax
\typeout{** WARNING: IEEEtran.bst: No hyphenation pattern has been}%
\typeout{** loaded for the language `#1'. Using the pattern for}%
\typeout{** the default language instead.}%
\else
\language=\csname l@#1\endcsname
\fi
#2}}
\providecommand{\BIBdecl}{\relax}
\BIBdecl

\bibitem{demene_spatiotemporal_2015}
C.~{Demené} \emph{et~al.}, ``Spatiotemporal clutter filtering of ultrafast
  ultrasound data highly increases doppler and fultrasound sensitivity,''
  \emph{IEEE Trans. Med. Imag.}, vol.~34, no.~11, pp. 2271--2285, Nov. 2015.

\bibitem{Wright_robust_2009}
J.~Wright \emph{et~al.}, ``Robust principal component analysis: Exact recovery
  of corrupted low-rank matrices via convex optimization,'' \emph{Neural Inf.
  Process. Syst}, 2009.

\bibitem{Fatemi2018}
M.~Bayat and M.~Fatemi, ``Concurrent clutter and noise suppression via low rank
  plus sparse optimization for non-contrast ultrasound flow doppler processing
  in microvasculature,'' in \emph{IEEE Int. Conf. Acoust., Speech Signal
  Process. (ICASSP)}, Calgary, Canada, April 2018.

\bibitem{Shen2019}
H.~{Shen} \emph{et~al.}, ``High-resolution and high-sensitivity blood flow
  estimation using optimization approaches with application to vascularization
  imaging,'' in \emph{EEE Int. Ultrason. Symp. (IUS)}, 2019, pp. 467--470.

\bibitem{Pham2020}
D.~H. {Pham}, A.~{Basarab}, I.~{Zemmoura}, J.~P. {Remenieras}, and
  D.~{Kouamé}, ``Joint blind deconvolution and robust principal component
  analysis for blood flow estimation in medical ultrasound imaging,''
  \emph{IEEE Trans. Ultrason., Ferroelect., Freq. Control}, pp. 1--1, 2020.

\bibitem{Taxt1995}
T.~{Taxt}, ``Restoration of medical ultrasound images using two-dimensional
  homomorphic deconvolution,'' \emph{IEEE Trans. Ultrason., Ferroelect., Freq.
  Control}, vol.~42, no.~4, pp. 543--554, 1995.

\bibitem{Michailovich2019}
O.~{Michailovich} \emph{et~al.}, ``Iterative reconstruction of medical
  ultrasound images using spectrally constrained phase updates,'' in \emph{IEEE
  16th Int. Symp. Biomed. Imag. (ISBI)}, April 2019, pp. 1765--1768.

\bibitem{Boyd_distributed_2010}
S.~Boyd \emph{et~al.}, ``Distributed optimization and statistical learning via
  the alternating direction method of multipliers,'' \emph{Found. Trends Mach.
  Learn.}, vol.~3, no.~1, pp. 1--122, 2010.

\bibitem{Cai2010}
J.-F. Cai, E.~J. Cand{\`{e}}s, and Z.~Shen, ``A singular value thresholding
  algorithm for matrix completion,'' \emph{SIAM J. Optim.}, vol.~20, no.~4, pp.
  1956--1982, Jan. 2010.

\bibitem{Keshavan2010}
R.~H. Keshavan, A.~Montanari, and S.~Oh, ``Matrix completion from noisy
  entries,'' \emph{J. Mach. Learn. Res.}, vol.~11, p. 2057–2078, August 2010.

\bibitem{Rodriguez2013}
P.~{Rodriguez} and B.~{Wohlberg}, ``Fast principal component pursuit via
  alternating minimization,'' in \emph{2013 IEEE Int. Conf. on Image Process.},
  2013, pp. 69--73.

\bibitem{Larsen1998}
R.~M. Larsen, ``Lanczos bidiagonalization with partial reorthogonalization,''
  \emph{{DAIMI} Report Series}, vol.~27, no. 537, Dec. 1998.

\bibitem{Larsen2005}
------, ``Propack,'' available at:
  \url{http://sun.stanford.edu/~rmunk/PROPACK/}, March 2004.

\bibitem{Tianyi2011}
T.~Zhou and D.~Tao, ``Godec: Randomized low-rank \& sparse matrix decomposition
  in noisy case,'' ser. ICML'11.\hskip 1em plus 0.5em minus 0.4em\relax
  Madison, WI, USA: Omnipress, 2011.

\bibitem{RodriguezMolares2018}
A.~Rodriguez-Molares \emph{et~al.}, ``The generalized contrast-to-noise
  ratio,'' in \emph{IEEE Int. Ultrason. Symp. (IUS)}.\hskip 1em plus 0.5em
  minus 0.4em\relax {IEEE}, Oct. 2018.

\end{thebibliography}
%\bibliographystyle{IEEEbib}
%\bibliography{refs_ISBI20}

%
%IEEE-ISBI supports the standard requirements on the use of animal and
%human subjects for scientific and biomedical research. For all IEEE
%ISBI papers reporting data from studies involving human and/or
%animal subjects, formal review and approval, or formal review and
%waiver, by an appropriate institutional review board or ethics
%committee is required and should be stated in the papers. For those
%investigators whose Institutions do not have formal ethics review
%committees, the principles  outlined in the Helsinki Declaration of
%1975, as revised in 2000, should be followed.

\end{document}